\documentstyle[aps,prb,psfig]{revtex}
\begin{document}

\title{Two-band theory of specific heat and thermal conductivity in the mixed
state of  MgB$_2$}
\author{L. Tewordt and D. Fay}
\address{I. Institut f\"ur Theoretische Physik,
Universit\"at Hamburg, Jungiusstr. 9, 20355 Hamburg, 
Germany}
\date{\today}
\maketitle
\begin{abstract}
      We solve the coupled gap equations for the $\sigma$- and
$\pi$-bands of MgB$_2$ in the vortex state and calculate the
resulting field dependencies of the specific heat coefficient
$\gamma$ and the thermal conductivity $\kappa\,$. The
crucial parameters of the theory are the interband pairing 
interaction $\lambda_{\pi\sigma}$ and the ratio 
$s=\xi_{\sigma}/\xi_{\pi}$ of the coherence lengths. For
reasonably small $\lambda_{\pi\sigma}$ and s, the small 
gap $\Delta_{\pi}$ decreases with increasing magnetic 
field $H$ much faster than the large gap  $\Delta_{\sigma}\,$.
This gives rise to the observed rapid increase of 
$\gamma_{\pi}$ and $\kappa_{\pi}$ for small fields while
$\gamma_{\sigma}$ and $\kappa_{\sigma}$ exhibit
conventional field dependencies. Inclusion of intraband
impurity scattering yields fairly good agreement with 
experiments for applied fields along the c axis.
\end{abstract}
\pacs{74.20.Fg, 74.70.Ad, 74.25.Op, 74.25.Fy}

         Evidence for the for the existence of two superconducting gaps in
 MgB$_2\,$ \cite{Nagamatsu} is provided by the rapid rise of the specific
heat coefficient $\gamma_s(H)\,$ \cite{Bouquet}  and the thermal
conductivity $\kappa_s(H)$ \cite{Solog} at very low fields. These measured
field dependencies can be explained qualitatively by assuming two 
independent bands where the large s-wave pairing gap 
$\Delta_{\sigma}\equiv\Delta_1$ is associated with the two-dimensional 
$\sigma$-band and the small s-wave gap $\Delta_{\pi}\equiv\Delta_2$ is
associated with the three-dimensional  $\pi$-band. \cite{TF1}  The steep rise
of $\gamma_s(H)$ and $\kappa_s(H)$ can be explained qualitatively by 
assuming that the "virtual" upper critical field for $\Delta_{\pi}$ is much 
smaller than that of $\Delta_{\sigma}\,$. \cite{Bouquet}  In the present paper
we improve the theory of Ref.~\onlinecite{TF1} by taking into account the 
interband pairing interaction while neglecting the interband impurity
scattering \cite{SchopScharn} which has been shown to be small.
\cite{Mazinetal}  We first have to solve the two-band
gap equations in the presence of the vortex lattices produced by a magnetic
field. Generalization of the linearized gap equations near the upper critical 
field $H_{c2}\,$ \cite{DahmSchop} to all averaged fields $H$ between 
$H_{c2}$ and $0$ yields, instead of the single gap equation of 
Ref.~\onlinecite{TF1}, the following coupled gap equations for the gaps 
$\Delta_i$ at $T=0\,$:
\begin{equation}
\Delta_i=\sum_{j=1}^2 \lambda_{ij} \int^{\omega_c}_0 
d\omega B_j(\omega,\Delta_j,\Lambda/v_j)\, . 
\qquad (i=1,2)
\label{GapGl1}
\end{equation}
Here, the $\lambda_{ij}$ are the intra- and interband pairing 
interactions multiplied by the densities of states $N_i(0)\,$, and the $B_j$ 
are the spectral functions of the anomalous propagators for the Abrikosov
vortex lattice: \cite{TF1}
\begin{equation}
B_i = \mbox{Re} \left[
\frac{-i\,\sqrt{\pi}\,\,2\,\Delta_{ri}\,(\Lambda/v_i)\,w(z_i)}
{ \left\{ 1 + 8\Delta^2_{ri}(\Lambda/v_i)^2[1+i\sqrt{\pi}\,z_i\, w(z_i)]
\right\}^{1/2} } \right]\, ,
\label{B}
\end{equation}
\begin{equation}
z_i = 2(\omega + i\gamma_i)\,\Lambda/v_i\, ;
\quad \Lambda=(2eH)^{-1/2}\, ;\quad \gamma_i = \Gamma_iA_i\, ;
\label{z}
\end{equation}
\begin{equation}
\Delta_{ri} = \Delta_i/D_i\, ;
\quad D_i = 1 - 2\, (\gamma_i \Lambda/v_i)\,w(z_i)\, .
\label{Del}
\end{equation}
The $\Gamma_i$ are the normal state impurity scattering rates, the
$A_i$ are the field-dependent densities of states, the $v_i$ are the
Fermi velocities perperdicular to the field, and the function $w(z)$ is
defined in Ref.~\onlinecite{TF1}. For band 2 (the $\pi$-band)
we assume a spherical Fermi surface. Then $v_2$ is replaced by
$v_2\sin\theta$ and $B_2$ and $D_2$ are averaged over the polar angle 
$\theta$ with respect to the direction of $H\,$. For brevity we omit here, and 
in the following, the terms containing $\sin\theta$ and the integrations 
over $d\theta\,\sin\theta$ from 0 to $\pi/2\,$. For band 1 (the $\sigma$-band) 
and the field along the c axis, $\theta=\pi/2$ and thus $\sin\theta=1\,$.

       In the limit $H\rightarrow0$ the vortex lattice constant 
$a=(2/\pi)^{1/2}\,\Lambda$  tends to infinity. Making use of the asymptotic 
expansion $w(z)\sim i/\sqrt{\pi}\,\,z\,$, the gap equations, Eq.(\ref{GapGl1}), 
become
\begin{equation}
\Delta_{i0}=\sum_{j} \lambda_{ij} \int^{\omega_c}_0 
d\omega\, \mbox{Re}\left[\frac{\Delta_{j0}}
{\left( \omega^2 - \Delta^2_{j0}\right)^{1/2}}\right]\, , 
\label{GapGl2}
\end{equation}
where the $\Delta_{i0}$ are the gap values at $T=0$ in zero field. Here we
have made use of the relation
\begin{equation}
(\omega + i\gamma_{i0})/\Delta^0_{ri}=
(\omega + i\gamma_{i0})\,D_{i0}/\Delta_{i0}=\omega/\Delta_{i0}
\label{omega}
\end{equation}
which can be derived with the help of the expression for $D_i$ in 
Eq.(\ref{Del}). However, this relation only holds in the absence of
interband impurity scattering. \cite{SchopScharn}

     With the help of the Abrikosov parameters, denoted by $\beta_i\,$,
we now express $\Lambda/v_i$ in terms of the reduced field $h=H/H_{c2}$ 
and the zero-field gap $\Delta_{i0}\,$:\, \cite{TF1}
\begin{equation}
\Lambda/v_1 = (6\beta_1\,h\,\Delta^2_{10})^{-1/2}\,;
\quad \Lambda/v_2 =s\, (6\beta_2\,h\,\Delta^2_{20})^{-1/2}\,;
\quad s=(v_1/v_2)(\Delta_{20}/\Delta_{10})=\xi_{10}/\xi_{20}\, .
\label{defs1}
\end{equation}
Employing these relations we can express the gaps $\Delta_i$ and the
scattering rates $\Gamma_i$ in Eq.(\ref{B}) by their ratios with respect to 
the $\Delta_{i0}\,$, and we can convert the integrations over $\omega$ in
Eq.(\ref{GapGl1}) to integrations over the new variables 
$\Omega= \omega/\Delta_{i0}\,$. We then divide Eq.(\ref{GapGl1}) by
$\Delta_i$ and Eq.(\ref{GapGl2}) by $\Delta_{i0}$ and subtract the latter
from the former. In this way we obtain two coupled equations for the two
unknown functions $x_1(h)$ and $x_2(h)$ for given values of the parameters
$\lambda_{ij}\,,\;\delta_i\,,\;\beta_i\,,\;r\,$, and $s\,$.   The quantities 
$x_i\,,\;\delta_i\,$, and $r$ are defined by
\begin{equation}
x_i = \Delta_i/\Delta_{i0}\, ; \quad \delta_i=\Gamma_i/\Delta_{i0}\,;
\quad r=\Delta_{20}/\Delta_{10}\, .
\label{defs2}
\end{equation}
In Fig. (1) we show the reduced gap functions $x_1(h)$ and $x_2(h)$ for
3 sets of parameter values:  I) $\lambda_{11}=1\,$, $\lambda_{22}=0.28\,$, 
$\lambda_{21}=0.17\,$, $\lambda_{12}=0.23\,$, $\beta_1=1.15\,$,
$\beta_2=1.58\,$, $s=1/6\,$; Ia) $\lambda_{ij}$ as in set I, $s=1/4\,$,
$\beta_1=1.21\,$, $\beta_2=1.58\,$; II)  $\lambda_{11}=1\,$, 
$\lambda_{22}=0.45\,$, $\lambda_{21}=0.16\,$, $\lambda_{12}=0.21\,$, 
$\beta_1=1.16\,$, $\beta_2=1.58\,$, $s=1/6\,$. The gap ratio has been taken
to be $r=1/3\,$ and the reduced impurity scattering rates 
$\delta_i = \Gamma_i/\Delta_{i0}$ are $\delta_1 = 0.5$ and $\delta_2 = 0.8\,$.
The $\lambda$-matrix elements in I) and II) have been obtained from band
structure calculations (Refs.~\onlinecite{Liu} and ~\onlinecite{Goletal}, see the 
review, Ref.~\onlinecite{MazinAnt}). The ratio of Fermi velocities is about 
$v_1/v_2 = 0.54\,$, \cite{Brink} and the gap ratio $r$ ranges between about
1/3 and 0.44\, \cite{Esk,Iav} which yields a range of the ratio $s$ of coherence
lengths (see Eq.(\ref{defs1}))
 between 0.18 and 0.24.
While the function $x_1(h)\simeq(1-h)^{1/2}$ is rather independent of the choice
of parameters, the function $x_2(h)$ depends sensitively on the values of 
$\lambda_{ij}$ and $s\,$. For vanishing interband coupling 
$\lambda_{21}$ we obtain approximately 
$x_2(h)\simeq(1-h/s^2)^{1/2}$ which goes to zero at a smaller effective upper
critical field $H^{\pi}_{c2}= s^2H_{c2}$ where $s\simeq\xi_{10}/\xi_{20}$ 
(see Eq.(\ref{defs1})). Thus we see that $H^{\pi}_{c2}$ corresponds to the "virtual" 
upper critical field for the $\pi$-band which was introduced in 
Ref.~\onlinecite{Bouquet} as the field above which the overlap of the vortex 
cores with large radius $\xi_{20}$ (see the STS measurements of 
Ref.~\onlinecite{Esk}\,) drives the majority of the ${\pi}$-band electrons normal.

     In Fig.(2) we have plotted our results for the zero energy density of states 
$A_i(\omega=0)$ obtained from the expression for $A_i$ (given by 
Eq.(\ref{B}) with the numerator set equal to 1) by inserting the previously 
calculated gap ratios $x_i(h)$ together with Eqs.(\ref{defs1}) for the functions 
$\Lambda(h)/v_i\,$. $A_2$ is obtained by averaging $A_2(\theta)$ over the
polar angle $\theta\,$. We note that it is important to calculate the impurity 
scattering rates $\gamma_i=\Gamma_i\,A_i(h)$ self-consistently. One sees 
from Fig. (2) that $A_2(h)$ rises steeply for small fields $h$ and then becomes
almost constant above $h\sim0.2\,$. The slope at $h=0$ and the downward
curvature for low fields increase as $s$ is decreased from 1/4 to 1/6. 
The function $A_1(h)$ is very similar to the function obtained previously for
a single band. \cite{TF1} The initial steep rise of $A_2(h)$ for $s=1/6$ qualitatively
fits the data points for the contribution of the $\pi$-band to the specific heat 
coefficient $\gamma(H)\,$.  \cite{Bouquet} The function $A_1(h)$ corresponds 
to the straight line assumed in Ref.~\onlinecite{Bouquet} for the $\sigma$-band 
contribution to $\gamma(H)$ for fields applied along the c axis.

      We turn now to the calculation of the in-plane electronic thermal
conductivity $\kappa_s(h)$ which is given at $T=0$ by the expression in
Ref.~\onlinecite{TF2}. Again it is important to take into account the 
renormalization of the gap by the function $D_i$ in the presence of 
impurity scattering (see Eq.(13) of Ref.~\onlinecite{TF1}). By inserting the
functions $x_i(h)$ obtained from the gap equations, Eq.(\ref{GapGl1}), and 
the functions $\Lambda(h)/v_i$ from Eq.(\ref{defs1}) into the expresions for
the ratios $\kappa_s/\kappa_n$ we obtain, for applied fields
\boldmath $H$ \unboldmath along the c axis, the thermal conductivity ratios 
$\kappa_{si}(h)/\kappa_{ni}$ shown in Fig.3. It should be noted that the 
$\pi$-band contribution $\kappa_{s2}(h)/\kappa_{n2}$ has been obtained 
as an average over the polar angle $\theta$ by including the factor 
$(3/2)\sin^2\theta$ which arises from the square of the group velocity in 
the ab plane. The curve for the $\sigma$-band conductivity
$\kappa_{s1}(h)/\kappa_{n1}$ turns out to be very similar to the curve obtained
in  Ref.~\onlinecite{TF1} for a single band with the same impurity scattering
rate $\delta_1=0.5\,$. The curve $\kappa_{s2}(h)/\kappa_{n2}$ for the 
$\pi$-band contribution with $\delta_2=0.8$ rises almost linearly with $h$ where
the slope near $h=0$ and the downward curvature for low fields increase as $s$
is decreased from 1/4 to 1/6. For applied fields perpendicular to the 
c axis, the measured thermal conductivity $\kappa_e$ first rises steeply for
small fields and then saturates while, for fields along the c axis, it exhibits an
upward curvature towards $H_{c2}\,$. \cite{Solog} These different behaviors
have been explained by separating the individual contributions of the $\pi$- and
$\sigma$-bands. Then $\kappa_{\pi}$ rises steeply with $H$ and 
approximately attains its normal-state value at a small field and 
$\kappa_{\sigma}$ first rises very slowly and then curves upward towards
$H_{c2}\,$.  \cite{Solog} These experimental curves are similar to our results
for $s=1/6$ shown in Fig. (3).

      We now briefly discuss the parameter values and approximations that have
been used to derive our results. We have seen that the rapid increase of the
specific heat coefficient $\gamma_s(h)/\gamma_n$ and the thermal conductivity
ratio $\kappa_s(h)/\kappa_n$ with increasing field $h=H/H_{c2}$ is due 
mainly to the $\pi$-band contribution. The reason is that the gap $\Delta_2(h)$ 
associated with the $\pi$-band almost closes at the so-called "virtual" upper
critical field\, \cite{Bouquet} $\,H^{\pi}_{c2}\sim s^2H_{c2}$ because the ratio 
$s=\xi_{10}/\xi_{20}$ of the coherence lengths of the $\sigma$- and the 
$\pi$-bands is much smaller than 1. For the most important parameters
entering our gap equations, $s$ and $r=\Delta_{20}/\Delta_{10}\,$, we have 
used the values $r=1/3$ and $s=$ 1/4, and 1/6 which are based on
various experiments on MgB$_2\,$. \cite{Bouquet,Brink,Esk,Iav} The field dependence
of the gap ratio $x_1(h)=\Delta_1/\Delta_{10}\,$, and thus of the contributions to
$\gamma_{s1}/\gamma_{n1}$ and $\kappa_{s1}/\kappa_{n1}$ arising from the 
$\sigma$-band, are nearly the same as those obtained for an independent single
$\sigma$-band, \cite{TF1} indicating that the effect of the interband coupling
$\lambda_{12}$ is rather small. However, the field dependence of 
$x_2(h)=\Delta_2/\Delta_{20}$ differs substantially from that for the independent single
$\pi$-band with an effective upper critical field $h^{\pi}_{c2}=s^2$ as can be seen from
Fig.(1). This is because $x_2(h)$ is non-zero between $h^{\pi}_{c2}$ and $h=1$ due to
the effect of the interband coupling $\lambda_{21}\,$. This shows that
superconductivity survives even though the vortex cores for the $\pi$-band with 
giant radius $\xi_{20}\,$ \cite{Esk} start to overlap for $h>h^{\pi}_{c2}\,$. As can be
seen in Fig.(1), the curve for $x_2(h)$ is very sensitive to the values of 
$\lambda_{ij}$ and $s\,$. We find that the experimental contributions to $\gamma$
and $\kappa$ arising from the $\pi$-band \cite{Bouquet,Solog} can be fitted by
taking the $\lambda_{ij}$ given by band structure calculations. \cite{Liu,Goletal,MazinAnt} 
The other crucial parameter values needed to obtain good fits of the data are
$r=1/3$ and $s=1/6$ which lie in the ranges obtained from experiment. The other
parameter values used in our numerical calculations are the reduced impurity
scattering rates $\delta_1=0.5$ and $\delta_2=0.8$ which have been estimated
from the relevant experiments. It turns out that even for these moderately large
impurity scattering rates it is very important to take into account the renormalization
of the gap (see Eq.(\ref{Del})) which leads to a large reduction of the effect of 
impurity scattering in comparison to that calculated without the function $D\,$. It is
also important that the calculation of the scattering rate 
$\gamma_i=\Gamma_i\,A_i(h)$ in the Born limit be carried out self-consistently 
together with the calculation of the zero-energy density of states $A_i(h)\,$. This
yields $A_i(0)=0\,$, as it should. The shape of the Fermi surface (FS) and the 
direction of the applied field play an important role because the spectral functions
$B_i$ in Eq.(\ref{B}) have to be averaged over the corresponding FS where the
velocity $v_i$ denotes the component $v_{i\perp}(\mathbf{p})\,$ perpendicular to
\boldmath $H\,$. \unboldmath For the spherical FS we have assumed for the $\pi$-band,
$v_{\perp}(\mathbf{p})=v\sin\theta\,$, where 
$\theta=\angle(\mathbf{p},$\boldmath$H$\unboldmath$)\,$.
In the limit $\theta\rightarrow0$ the function $B$ approaches the BCS spectral 
function of the anomalous propagator. We find that the results for the averages over the 
polar angle $\theta$ do not differ significantly from the results obtained by setting
$\theta=\pi/2\,$. This means that the quasiparticles moving perpendicular to the 
vortex axes yield the dominant contributions to $\gamma$ and $\kappa\,$. We have 
approximated the $\pi$-band FS by a sphere whereas the $\lambda_{ij}$ have been
calculated for the actual FS. This actual FS can be modeled by a half-torus 
\cite{DahmSchop} which yields, with the $\lambda$-matrix of Ref.~\onlinecite{Liu} and
small $s$, results \cite{DahmGlSh} which agree qualitatively with ours shown in Figs.
1 and 2 for $s=1/6\,$. Finally it should be pointed out that we have employed the 
Abrikosov ground state of the vortex lattice although, in particular at lower fields, a 
Landau-level expansion or a variational expression\, \cite{DahmSchop} is needed to 
describe the distorted vortex lattice. The results for $\gamma_s(h)/\gamma_n$ and 
$\kappa_s(h)/\kappa_n$ shown in Figs.(2) and (3) for the $\sigma$- and $\pi$-bands 
should still be added by weighting them with the corresponding density of states. 

      In conclusion we can say that our two-band theory for the vortex state in 
MgB$_2$ can satisfactorily account for the observed field dependence of the 
specific heat coefficient $\gamma$ and the thermal conductivity $\kappa\,$. The
small gap $\Delta_2$ associated with the $\pi$-band decreases with increasing field
$H$ much faster than the large $\sigma$-band gap $\Delta_1$ which shows 
conventional field dependence. This gives rise to the rapid increase of $\gamma$
and $\kappa$ at small fields. Due to a small interband pairing interaction 
$\lambda_{21}\,$, the gap $\Delta_2$ remains finite even in the field region where the
large $\pi$-band vortex cores of radius $\xi_{20}$ overlap. This leads to smooth 
evolution of the $\pi$-band contributions to $\gamma$ and $\kappa$ to their normal
state values near a "virtual" upper critical field $H^{\pi}_{c2} \simeq s^2H_{c2}$ which is
much smaller than $H_{c2}$ because the ratio $s=\xi_{10}/\xi_{20}$ is much smaller
than 1. The Fermi surface topology and the impurity scattering have relatively small
influence on the field dependence of $\gamma$ and $\kappa$ for fields applied
along the c axis.

\bigskip
We thank T. Dahm for valuable discussions.
\newpage
\newpage
\begin{figure}
\centerline{\psfig{file=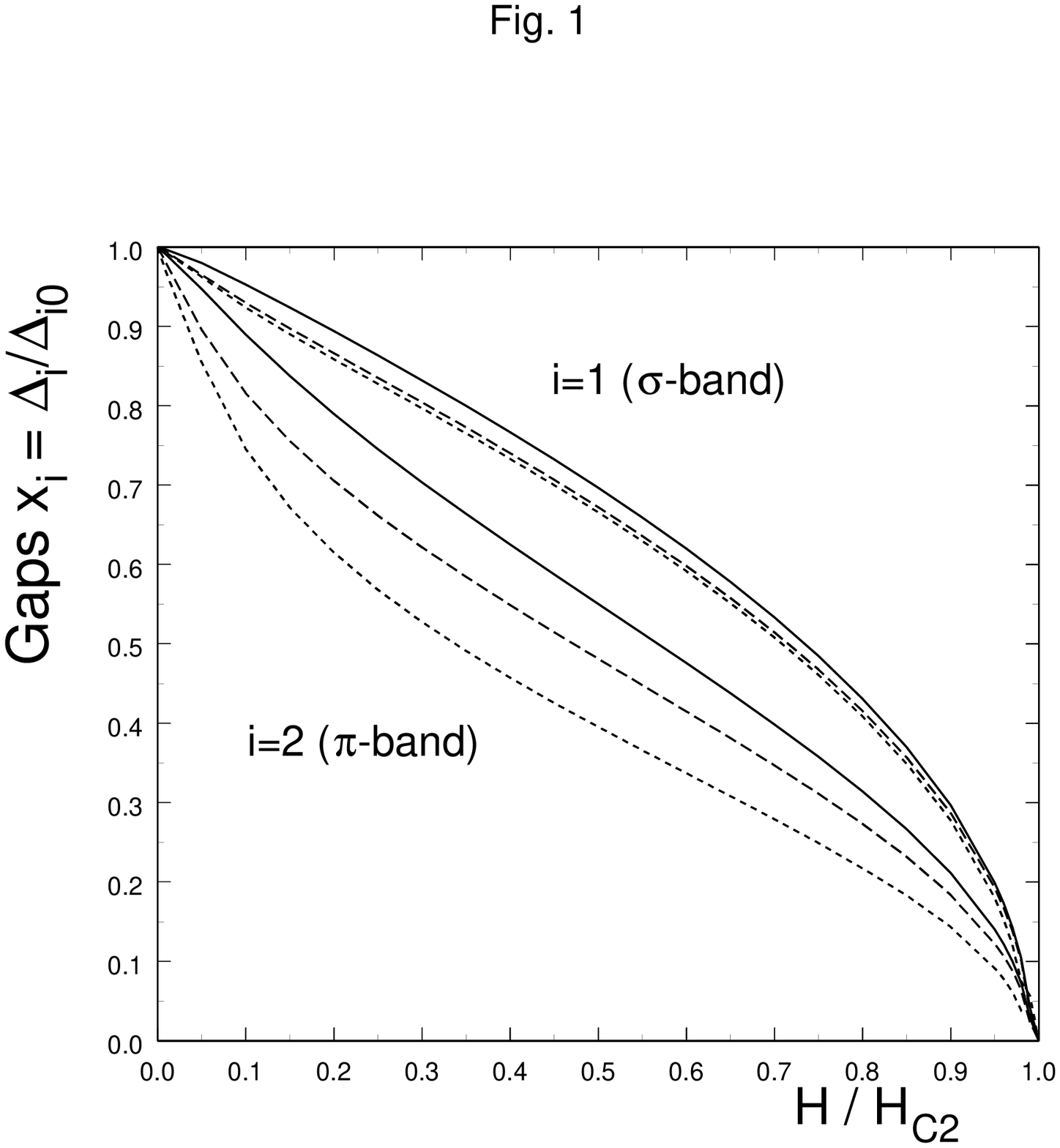,width=18cm,angle=0}}
\vskip -2cm
\caption{Reduced gaps $x_1(h)=\Delta_1/\Delta_{10}$ for the $\sigma$-band and 
$x_2(h)=\Delta_2/\Delta_{20}$ for the $\pi$-band vs reduced magnetic field
$h=H/H_{c2}$ for applied fields along the c axis. The upper curves are for $x_1$
and the lower curves for $x_2\,$. The pairing interaction matrices $\lambda_{ij}$
have the values given in Ref. 8 (our parameter sets I, $s=1/6$, and 
Ia, $s=1/4$) and in Ref. 9 (set II, $s=1/6$) where $s$ is the 
ratio of coherence lengths $s=\xi_{10}/\xi_{20}\,$. The curves correspond to the
parameter sets Ia, I, and II, from top to bottom. The reduced impurity scattering 
rates $\delta_i$ are $\delta_1=0.5$  for the $\sigma$-band and $\delta_2=0.8$ 
for the $\pi$-band.}
\label{fig1}
\end{figure}
\begin{figure}
\centerline{\psfig{file=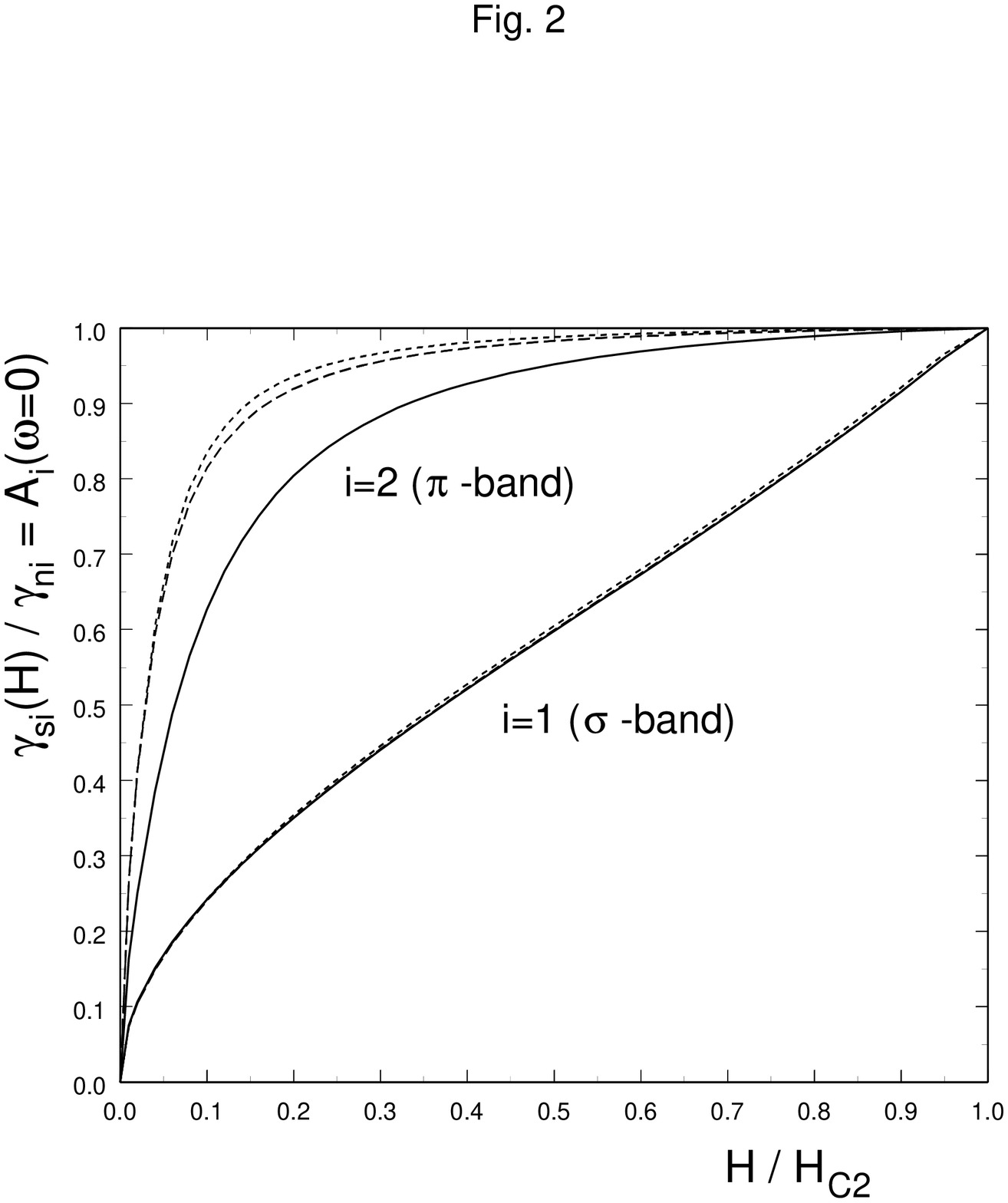,width=18cm,angle=0}}
\vskip -1cm
\caption{Specific heat coefficients, or zero-energy densities of states, 
$\gamma_{si}/\gamma_{ni}=A_i\,$, ($i=1,2$) vs $h$ for the parameter values of
Fig.(1). The lower curves are for the $\sigma$-band, $i=1\,$, and the upper curves 
are for the $\pi$-band, $i=2\,$, for parameter sets Ia, I, and II, from bottom to top.}
\label{fig2}
\end{figure}
\begin{figure}
\centerline{\psfig{file=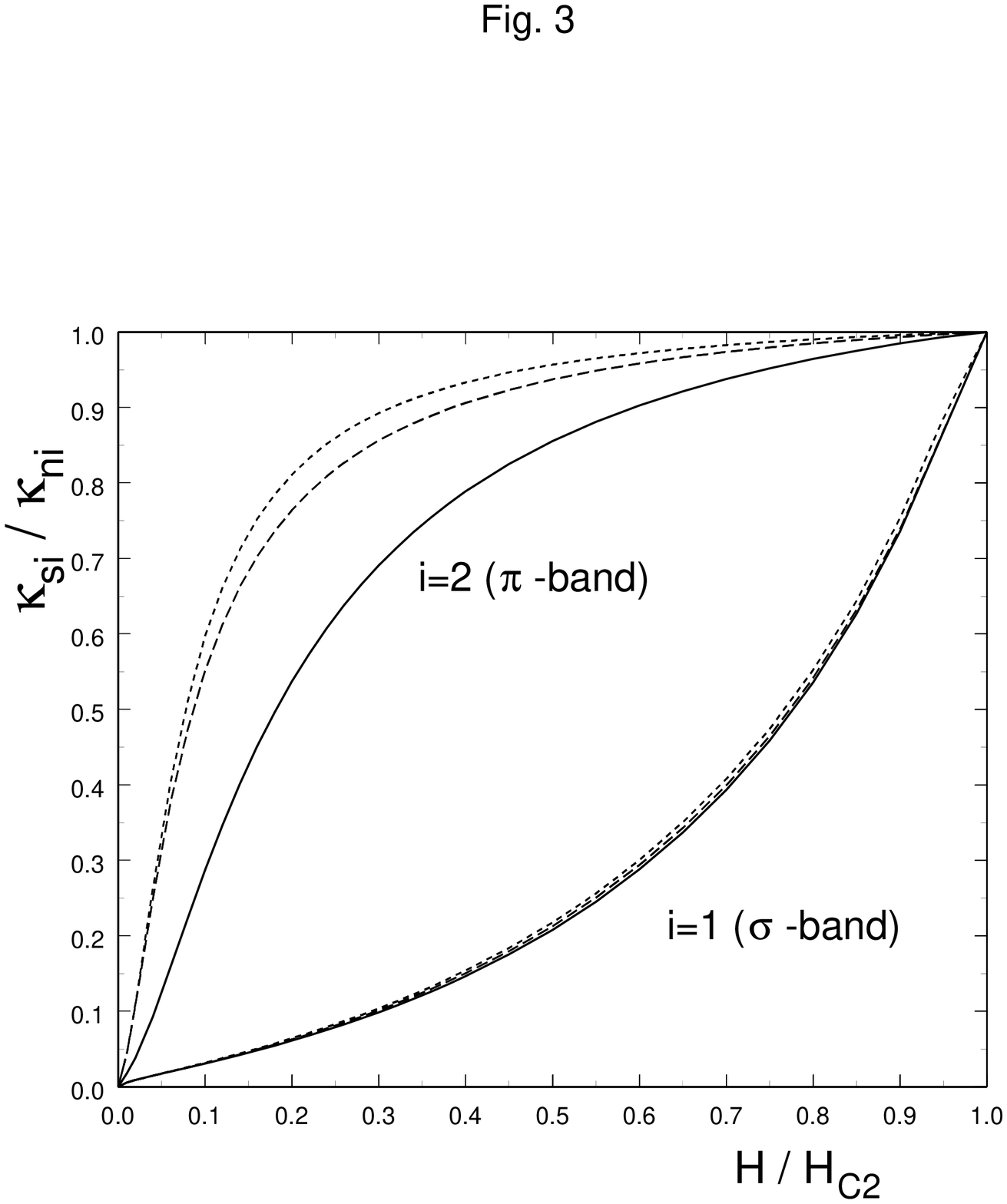,width=18cm,angle=0}}
\vskip -1cm
\caption{Reduced electronic thermal conductivities, $\kappa_{si}/\kappa_{ni}\,$, ($i=1,2$)
vs $h$ for the parameter values of Fig.(1). The lower curves are for the
$\sigma$-band, $i=1\,$, and the upper curves are for the $\pi$-band, $i=2\,$, for 
parameter sets Ia, I, and II, from bottom to top.}
\label{fig3}
\end{figure}

\begin{references}
%
%
\bibitem{Nagamatsu}J. Nagamatsu, N. Nakagawa, T. Muranaka,Y. Zenitani, 
and J. Akimitsu, Nature {\bf 410}, 63 (2001).
%
\bibitem{Bouquet}F. Bouquet, Y. Wang, I. Sheikin, T. Plackowski, A. Junod,
S. Lee, and S. Tajima,  Phys. Rev. Lett.  {\bf 89}, 257001-1 (2002).
%
\bibitem{Solog}A. V. Sologubenko, J. Jun, S. M. Kazakov, 
J. Karpinski, and H. R. Ott, Phys. Rev. B {\bf 66}, 014504 (2002).
%
\bibitem{TF1}L. Tewordt and D. Fay, Phys. Rev. B {\bf 67}, 134524 (2003).
%
\bibitem{SchopScharn}N. Schopohl and K. Scharnberg, Solid State
Commun. {\bf 22}, 371 (1977).
%
\bibitem{Mazinetal}I. I. Mazin et al., Phys. Rev. Lett. {\bf 89}, 107002 (2002).
%
\bibitem{DahmSchop}T. Dahm and N. Schopohl, cond-mat/0212188; Phys. Rev.
Lett. (in press).
%
\bibitem{Liu}A. Y. Liu, I. I. Mazin, and J. Kortus,  Phys. Rev. Lett.  {\bf 87}, 
087005 (2001).
%
\bibitem{Goletal}A. A. Golubov et al., J. Phys. Condens. Matter {\bf 14}, 
1353 (2002).
%
\bibitem{MazinAnt}I. I. Mazin and V. P. Antropov, Physica C {\bf 385}, 
49 (2003).
%
\bibitem{Brink}A. Brinkman et al, Phys. Rev. B {\bf 65}, 
180517(R) (2002).
%
\bibitem{Esk}M. R. Eskildsen, M. Kugler, S. Tanaka, J. Jun, S. M. Kazakov,
J. Karpinski, and \O. Fischer,  Phys. Rev. Lett. {\bf 89}, 187003-1 (2002).
%
\bibitem{Iav}M. Iavarone et al., Phys. Rev. Lett. {\bf 89}, 187002 (2002).
%
\bibitem{TF2}L. Tewordt and D. Fay, Phys. Rev. B {\bf 64}, 
24528 (2001).
%
\bibitem{DahmGlSh}T. Dahm, S. Graser, and N. Schopohl, Physica C
(Proceedings of the M2S-Rio, in press).
%
\end{references}
\end{document}